\documentclass[]{fairmeta}

\DeclareMathOperator*{\argmin}{arg\,min}

\title{Exploring Forum Post Retrieval with Generative Modeling}

\author[1,*, \dagger]{Yang Li}
\author[2, \dagger]{Yaguang Liu}
\author[2, \dagger]{Heng Liu}
\author[2, \dagger]{Samson Komo}
\author[2]{Jane Kou}
\author[2]{Yulian Zhou}
\author[2]{Gang Yang}
\author[2]{Shubhojeet Sarkar}
\author[2, \dagger]{Gaurav Chakravorty}
\author[2]{Yujie Liu}
\author[1]{Haipeng Chen}
\author[2]{Yonghuan Yang}
\author[2]{Deepti Chheda}
\author[2]{Yamin Wang}
\author[2]{Mike Plumpe}
\author[2]{Rish Tandon}
\author[2, \dagger]{Shengbo Guo}

\affiliation[1]{William \& Mary}
\affiliation[2]{Meta}

\contribution[*]{Work done during internship at Meta}
\contribution[\dagger]{Equal contribution.}

\abstract{Generative recommendation (GR) has emerged as an alternative to embedding-based retrieval, building on the success of generative models in language and vision. We are exploring GR on Facebook Forum, a standalone application for medium-to-heavy users of Facebook Groups. Because Forum is a new surface, its own interaction data are too sparse to train a GR model from scratch. We address this with transfer along two axes: we train on a broader corpus of Facebook Groups engagements rather than Forum sessions alone, and we reuse hierarchical, prefix-based semantic IDs (SIDs) learned from cross-platform Facebook Feed data instead of fitting a Forum-specific tokenizer. A 3B-parameter instruction-tuned language model is then supervised-fine-tuned to generate SIDs directly from user context. We systematically ablate the design choices that matter most in practice, including SID construction, the composition and length of user history, and the inclusion of user-profile features. Our results show that cross-platform SIDs transfer to a new recommendation surface, and offer practical guidance for teams deploying GR on real-world social platforms.}

\date{\today}
\correspondence{Yang Li \email{yli102@wm.edu}}

\usepackage[most]{tcolorbox}
\usepackage{xcolor}
\newcommand{\sidtok}[2]{\textit{\textless#1\_#2\textgreater}}
\usepackage[colorinlistoftodos]{todonotes}

\definecolor{promptframe}{RGB}{31,56,100}   
\definecolor{promptback}{RGB}{248,249,251}  

\newtcolorbox{promptbox}{
  colback=promptback,
  colframe=promptframe,
  boxrule=0.6pt,
  arc=6pt,                  
  left=8pt, right=8pt, top=6pt, bottom=6pt,
  fontupper=\small,
  breakable
}

\begin{document}

\maketitle





\section{Introduction}
\label{section:intro}
Facebook is one of the largest social platforms in the world, serving billions of people
across an extremely wide range of interests. Facebook Groups is among the largest
contributors to that daily activity: rather than a single global feed, Groups organizes
content around communities of shared interest, so the recommendation problem is
inherently one of matching a user to content within and across the specific communities
they care about.

\textbf{Forum} is a new standalone companion application built for people who are heavily
invested in Groups. It offers direct access to group discussions along with tools
intended to deepen community engagement, and it is deliberately narrower than the main
Facebook application: its users arrive with community-oriented intent rather than to
browse a general feed. This makes Forum an attractive setting for recommendation
research, but also a difficult one. As a newly launched surface, Forum has accumulated
comparatively little of its own interaction data, so a model trained on Forum sessions
alone would be badly under-served --- precisely the cold-start regime in which
transferring representations and behavioral signal from a larger related surface is most
valuable.

With the success of generative models in natural language processing, computer vision,
and time-series modeling, the recommender-systems community has begun to explore
generative recommendation (GR) as an alternative to embedding-based retrieval. Rather
than embedding a query and performing approximate nearest-neighbor search over an item
index, a GR model \emph{generates} an identifier for the next item directly. Work in this
direction spans diffusion-based approaches to sequential
recommendation~\cite{diffurec,dreamrec,diffrec} and autoregressive approaches that represent
items as discrete token sequences and predict the next item with a
transformer~\cite{tiger,lc-rec,p5}. The autoregressive formulation is a natural fit for our
setting: it lets us reuse a pretrained language model, and --- because items are
identified by semantic ID (SID) rather than by an opaque index --- the same model can
consume natural-language context about a user alongside their interaction history.

This paper reports our experience applying autoregressive GR to Groups content, as a step
toward serving Forum. Our broader effort aims to enhance the conventional retrieve stack with a new foundation model that generates candidate group posts; here we focus on the generative retrieval half, and on the
data and training foundations that any such system requires. Concretely, we build the post-to-SID mapping for Groups posts, construct the alignment corpus that teaches a pretrained model a new SID vocabulary, assemble user-level engagement sequences in SID
form, and train and evaluate models end-to-end on production Groups data.

Two forms of transfer make this feasible despite Forum's limited data. First, rather than
training on Forum sessions alone, we train on the substantially larger corpus of Facebook
Groups engagements. Second, rather than fitting a Forum-specific tokenizer, we reuse
hierarchical, prefix-based SIDs learned by an RQ-VAE trained on cross-platform Facebook
Feed content, which lets a new surface inherit an item vocabulary it could not have
learned on its own.

Our contributions are as follows:
\begin{itemize}
  \item We describe an end-to-end GR system for a real social platform, built on a
    3B-parameter instruction-tuned language model that generates cross-platform SIDs from
    user engagement history (Section~\ref{sec:generation}).
  \item We systematically evaluate the design choices that most affect quality in
    practice --- SID depth, how engagement history is verbalized in the prompt, history
    length and whether it must be fixed, and the choice of base model --- finding that SID
    depth dominates all other factors and that the smallest base model we tried is also
    the most cost-efficient (Section~\ref{sec:experiments}).
\end{itemize}
\section{Related Work}
\label{sec:related_work}

\paragraph{Large Language Models for Recommendation.}
The integration of Large Language Models (LLMs) into recommendation systems has fundamentally shifted the paradigm from traditional embedding-based retrieval to generative recommendation \cite{lin2023survey, wu2023survey, wang2023generative}. Early efforts treated recommendation as a language processing task, mapping user histories and item metadata into natural language prompts \cite{p5, li2023text}. Subsequent work has explored instruction tuning to better align pre-trained language models with recommendation objectives \cite{bao2023tallrec, zhang2023recommendation, liao2023llara}, while others have investigated LLMs as interactive and explainable agents \cite{liu2023chatgpt, gao2023chat}. Unlike approaches that rely purely on lengthy textual descriptions of items, our work leverages an instruction-tuned LLM to generate compact, discrete semantic identifiers, bridging the gap between natural language reasoning and strict item retrieval.

\paragraph{Semantic IDs and Generative Retrieval.}
To sidestep the inefficiencies of generating full text for items, recent systems employ generative retrieval, where models predict discrete item identifiers directly \cite{dsi, nci}. Semantic IDs (SIDs), typically derived from hierarchical clustering or quantization of item embeddings, preserve semantic proximity: items with similar content share prefix tokens \cite{tiger, lc-rec, hua2023index}. While early generative recommendation approaches used standard VQ-VAEs \cite{vqvae, hou2023vqrec}, we adopt Residual-Quantized VAEs (RQ-VAE) \cite{rqvae} to achieve a compact, hierarchical codebook. This strict prefix structure is crucial for transferring item representations across domains, allowing our model to leverage cross-platform Facebook Feed data for the cold-start Forum surface \cite{wang2023zeroshot, fan2023recommender}.

\paragraph{Reinforcement Learning for Alignment.}
Aligning language models with human preferences via Reinforcement Learning from Human Feedback (RLHF) \cite{instructgpt, bai2022training} has become standard practice for general-purpose LLMs. Early foundations demonstrated that fine-tuning on human preferences could drastically improve text quality and helpfulness \cite{ziegler2019fine, stiennon2020learning}. In recommendation, policy optimization techniques like Proximal Policy Optimization (PPO) \cite{schulman2017proximal} have historically been adapted to optimize ranking metrics directly \cite{zhao2018deep}. More recently, methods like Generative Reward Policy Optimization (GRPO) \cite{grpo} and Direct Preference Optimization (DPO) \cite{rafailov2023dpo} have shown promise by eliminating the need for separate value models. However, as we demonstrate in our post-training experiments, applying these RL alignment techniques to short-sequence generative recommendation introduces severe reward sparsity, highlighting a critical gap between text-based reasoning alignment and item-based ranking alignment.
\section{System Design}

In this section we describe the core components of our implementation.
As in autoregressive generative models for images~\cite{vqvae,rqvae}, recommendation
generation proceeds in two stages: (1) encoding a candidate post into a sequence of
discrete semantic tokens, and (2) learning a probabilistic model over these token
sequences.

\subsection{Semantic IDs via RQ-VAE}
\label{sec:sid}

The choice of encoding is one of the central design challenges in a real-world
deployment, since it determines both what the generative model can express and how
long its output sequences are. Continuous embeddings are the natural latent
representation for diffusion-based models, whereas codebook-based encodings such as
the Vector-Quantized Variational Autoencoder (VQ-VAE)~\cite{vqvae} yield the discrete
vocabulary that autoregressive transformers require.

The GR community has more recently adopted Residual-Quantized VAE
(RQ-VAE)~\cite{rqvae,tiger}, which reaches comparable reconstruction quality with a
codebook that is orders of magnitude smaller: rather than assigning one code per item
from a single large codebook, RQ-VAE represents an item as a short sequence of codes
that successively refine one another. This coarse-to-fine structure also gives the
resulting semantic IDs (SIDs) a prefix property that we exploit through experiment.

We adopt an internal RQ-VAE trained on Facebook Feed data, using post embeddings
produced by an internal content-understanding model. The model exposes two
configurations: a 3-layer and a 4-layer codebook, each layer holding $K = 2048$ codes.
We summarize the procedure below.

Let $C^l = \{c^l_k\}_{k=1}^{K}$ denote the codebook at layer $l \in [1, L]$, with
$|C^l| = K$. The encoder quantizes an item embedding $e$ into a sequence of $L$ codes,
and the decoder reconstructs $e$ from that sequence. Quantization is applied
recursively to the residual: setting $r^0 = e$,
\begin{equation}\label{eq:encoder}
  k_l = \argmin_{k \in [1,K]} \bigl\| r^{l-1} - c^l_k \bigr\|_2^2, \ \ \
  r^l = r^{l-1} - c^l_{k_l},
\end{equation}
so that the semantic ID of the item is the tuple
$\mathcal{RQ}(e) = (k_1, \dots, k_L)$ and the reconstruction is
$\hat{e} = \sum_{l=1}^{L} c^l_{k_l}$. Because each layer quantizes what the previous
layers left unexplained, the codes are ordered coarse-to-fine, and items sharing a
prefix are semantically related.

The codebooks, encoder, and decoder are trained jointly, minimizing a reconstruction
term together with the quantization error accumulated across layers:
\begin{equation}
  \mathcal{L}_{\text{total}}
  = \underbrace{\| e - \hat{e} \|_2^2}_{\mathcal{L}_{\text{recon}}}
  + \beta \sum_{l=1}^{L}
    \underbrace{\bigl\| \mathrm{sg}[r^{l-1}] - c^l_{k_l} \bigr\|_2^2
    + \bigl\| r^{l-1} - \mathrm{sg}[c^l_{k_l}] \bigr\|_2^2}_{\mathcal{L}_{\text{residual}}},
\end{equation}
where $\mathrm{sg}[\cdot]$ denotes the stop-gradient operator, which allows gradients
to pass through the non-differentiable $\argmin$ via the straight-through estimator.

\subsection{Generative Recommendation with a Small Language Model}
\label{sec:generation}

Following established practice in generative retrieval~\cite{tiger,lc-rec,dsi,nci}, we train a
small language model (SLM) to generate SIDs directly. Adapting a pretrained SLM to this
task requires two stages: (1) grounding the newly introduced SID tokens in the model's
existing textual knowledge, and (2) learning to generate the SID of a user's next
engagement from their interaction history.

\subsubsection{Align SIDs in Textual Context}
\label{sec:alignment}

The SLM arrives with strong instruction-following ability in the textual domain, but the
SID tokens are entirely new to it: an identifier such as \sidtok{a}{1029} is an
uninitialized embedding carrying no semantics. Our first stage bridges this gap, letting
the model reuse what it already knows about language to form useful representations of
each SID.

Finer-grained alignment supervision generally yields better representations, but
collecting it through human annotation does not scale. We instead exploit a pairing that
arises for free in a production setting: every post already has both a text body and an
SID assigned by the RQ-VAE of Section~\ref{sec:sid}. We use this pairing in both
directions, requiring no annotation:

\textbf{Text $\rightarrow$ SID} teaches the model to map a textual description onto the
identifier of the post it describes, aligning language directly with the SID vocabulary.
\begin{promptbox}
\textbf{Input:}\\ The content can be described as follows:
``Happy Mother's Day, Mom! I love you so much!''
What is its semantic ID? \\
\textbf{Response:}\\ \sidtok{a}{1029}\sidtok{b}{1520}\sidtok{c}{350}
\end{promptbox}

\textbf{SID $\rightarrow$ Text} teaches the inverse mapping, forcing each SID embedding
to become predictive of its post's content.
\begin{promptbox}
\textbf{Input:}\\ Describe the content associated with semantic ID
\sidtok{a}{1029}\sidtok{b}{1952}\sidtok{c}{861}. \\
\textbf{Response:}\\Happy Father's Day, Dad! I love you so much!
\end{promptbox}

Note that these two posts share the first-layer code \sidtok{a}{1029} while diverging at
the second: the coarse-to-fine structure of Section~\ref{sec:sid} places semantically
related content under a common prefix, and this stage exposes the SLM to exactly that
structure.

Because the goal here is representation learning rather than conditional generation, we
compute the language-modeling loss over all tokens in the sequence rather than masking the prompt as is standard for instruction tuning.
Gradients therefore reach the SID embeddings whether an SID appears as input or as
target.

\subsubsection{Learning to Generate Recommendations}

The second stage builds the recommendation capability itself. Given an instruction and a
fixed-length history of the user's recent engagements, each rendered as an SID, the model
is trained to generate the SID of the next post the user engages with positively.

The main difficulty is the composition of that history. We distinguish \emph{strong}
positive signals (like, share, comment, save) from \emph{weak} ones (dwell time above a
threshold $\tau$, group visit, profile visit). Strong signals are the prediction target,
but they are also sparse: a history assembled from strong signals alone spans a long and
uneven time window, so consecutive items in the sequence may be days apart and carry
little sequential relationship. We therefore populate the history with both weak and
strong signals, preserving chronological order, while supervising only on strong-signal
targets.

We explore two types of history composition: (1) pure SID history and (2) SID history with action names (i.e., liked \sidtok{a}{1}\sidtok{b}{1}\sidtok{c}{1}). An example of SID with action names is shown below:
\begin{promptbox}
\textbf{Input:} 

You are a helpful post recommendation specialist. Given a user's engagement history (e.g., reacted to, commented on, shared) along with each post's semantic ID (e.g., \sidtok{a}{1}\sidtok{b}{1}\sidtok{c}{1}), you will predict the semantic ID of the next post the user will positively engage with (i.e., like, share, comment, save, etc.).\\

The user has engagement history: viewed a photo \sidtok{a}{1}\sidtok{b}{1}\sidtok{c}{1},
viewed a video \sidtok{a}{2}\sidtok{b}{2}\sidtok{c}{2},
reacted to a photo \sidtok{a}{3}\sidtok{b}{3}\sidtok{c}{3}, \dots in chronological order. Based on this history, predict the next post the user will positively engage with (i.e., like, share, comment, save, etc.).\\
\textbf{Next item:} \\
\sidtok{a}{4}\sidtok{b}{4}\sidtok{c}{4}
\end{promptbox}

In contrast to the grounding stage, this stage follows the standard instruction-tuning
recipe: the loss is masked over the prompt so that gradients are computed only on the
response tokens and the history serves
purely as conditioning context. Concretely, with the prompt tokens masked out, the loss is
\begin{equation}
    \mathcal{L} = -\sum_{l=1}^{L} \log p_\theta(k_l \mid k_{<l}, \mathrm{prompt})
\end{equation}
i.e. standard next-token prediction over the target SID alone.

\section{Experiments}
\label{sec:experiments}

We study four questions that a practitioner must answer before deploying GR, and one that
determines its cost: (1) how deep should the SID hierarchy be? (2) how should the engagement
history be rendered into the prompt, and does an explicit alignment stage help? (3) how long
a history should the model condition on, and must that length be fixed? and (4) does the
choice of base model change any of these answers?

\subsection{Setup}

\textbf{Data and metrics.} Training sequences are drawn from Facebook Groups engagements
using the weak/strong signal taxonomy of Section~\ref{sec:generation}.
We report HR@$k$ for
$k \in \{3,5,10\}$ and NDCG@10 over a held-out set of 86{,}648 users. A hit requires the
generated SID to match the ground-truth item exactly in all $L$ layers.

\textbf{Matched-budget protocol.} All arms are trained to a fixed SFT step budget, and all
comparisons are drawn at the 200k-step checkpoint. Several arms were trained further, but we
restrict analysis to 200k so that every number in this section reflects the same budget.

\textbf{Reference configuration.} Unless stated otherwise, results use Llama-3.2-3B-Instruct
with 3-layer SIDs, a fixed history of 10 items, verbalized history, and CPT initialization.
This configuration scores HR@10 $=0.0667$ at 200k steps. Two conditions restrict the
eligible user population and are evaluated on smaller sets --- a 50-item history can only be
assembled for users with at least 50 engagements --- and we note the count wherever it
limits a comparison.

\subsection{SID Construction}
\label{sec:exp-sid}

\begin{table}[t]
\centering\small
\caption{SID depth and history rendering, HR@10 at 200k SFT steps, history length 10.
\emph{Verbs} prefixes each history item with its action
(``liked \sidtok{a}{10}\sidtok{b}{20}\sidtok{c}{30}''); \emph{raw SID} lists bare
identifiers.}
\label{tab:sid}
\begin{tabular}{llc}
\toprule
Depth & Rendering & HR@10 \\
\midrule
3-layer & Verbs (With Prompt Template)   & \textbf{0.0667} \\
3-layer & Raw SID (With Prompt Template)& 0.0654 \\
\addlinespace
4-layer & Verbs (With Prompt Template) & 0.0303 \\
4-layer & Raw SID (With Prompt Template) & 0.0296 \\
\bottomrule
\end{tabular}
\\[2pt]
\end{table}

SID depth is the single most consequential design choice we studied. At matched budget the
3-layer configuration reaches HR@10 of 0.0667 against 0.0303 for the 4-layer configuration
--- a $2.2\times$ difference, an order of magnitude larger than any other effect we measured.

We attribute this to task difficulty rather than to representation quality. Because a hit
requires every layer to be correct, a fourth layer imposes an additional 2048-way decision
on every prediction: with $K = 2048$, the nominal output space grows from
$2048^3 \approx 8.6 \times 10^9$ to $2048^4 \approx 1.8 \times 10^{13}$. The deeper SID is a
strictly more precise target, and the model pays for that precision in accuracy.

This makes the comparison a trade-off rather than a verdict. Coarser SIDs collide more: more
distinct posts map to the same code tuple, so a 3-layer hit identifies a semantic
neighborhood rather than a unique post, and HR@10 is correspondingly easier to satisfy. The
two configurations are therefore not scored against equally hard targets, and the
$2.2\times$ gap overstates the practical advantage by an amount related to the collision
rate.
Which depth to prefer thus depends on what sits downstream. If generation feeds a ranking
stage that can disambiguate within a colliding tuple, the coarser SID is the better
retrieval target; if the generated SID must identify a single post, the 4-layer cost may be
unavoidable.

\subsection{Prompt Construction and Continued Pre-Training}
\label{sec:exp-prompt}

Two components of our design aim at the same goal: letting the model bring its pretrained language understanding to bear on a vocabulary it has never seen. The alignment stage of Section~\ref{sec:alignment} does so before SFT, by aligning SID tokens with post text.
Prompt verbalization does so during SFT, by annotating each history item with the action that produced it. We disentangle the two in a $2 \times 2$ design.

\begin{table}[t]
\centering\small
\caption{Prompt rendering $\times$ continued pre-training, 200k SFT steps, 3-layer SIDs,
history length 10, all arms on the same 86{,}648-user evaluation set. \emph{CPT} initializes
from the text$\leftrightarrow$SID aligned checkpoint; \emph{stock} initializes from
Llama-3.2-3B-Instruct directly.}
\label{tab:prompt-cpt}
\begin{tabular}{llcccc}
\toprule
Init & Rendering & HR@3 & HR@5 & HR@10 & NDCG@10 \\
\midrule
CPT Model   & Verbs (With Prompt Template) & \textbf{0.0399} & \textbf{0.0509} & \textbf{0.0667} & \textbf{0.0420} \\
Stock Model& Verbs (With Prompt Template)  & 0.0391 & 0.0492 & 0.0640 & 0.0406 \\
CPT  Model & Raw SID (No Prompt Template) & 0.0383 & 0.0484 & 0.0633 & 0.0399 \\
Stock Model& Raw SID (No Prompt Template)& 0.0379 & 0.0479 & 0.0621 & 0.0393 \\
\bottomrule
\end{tabular}
\end{table}

Both interventions help, and they reinforce one another. Verbalizing the history improves HR@10 slightly by 2.0\% relative when the model has been grounded (0.0667 vs.\ 0.0654) for 3-layer SID and by 2.4\% relative for 4-layer SID. Symmetrically, continued pre-training improves HR@10
by 4.2\% with a verbalized prompt (0.0667 vs.\ 0.0640) and by 1.9\% with raw SIDs
(0.0633 vs.\ 0.0621). The two are complementary rather than redundant: each is worth roughly
twice as much in the presence of the other.

That interaction is the clearest evidence we have for the mechanism we intended. Grounding
teaches the model to relate SID tokens to language; verbalization gives it language in the
prompt to relate them to. Neither alone realizes the full benefit, and a model that had
reduced the task to transition statistics over the code vocabulary would show no interaction
at all --- it would be indifferent to natural-language annotations interleaved between SIDs
regardless of initialization.

The practical ordering is nonetheless worth stating plainly: the verbalized prompt
\emph{without} continued pre-training (0.0640) outperforms raw SIDs \emph{with} it (0.0633).
A change to the prompt template --- a few tokens per history item, no additional training ---
buys more than a full continued-pre-training stage over a bidirectional alignment corpus. A
team with a fixed budget should verbalize before it aligns, then align if budget remains. 

\subsection{History Length}
\label{sec:exp-history}

\begin{table}[t]
\centering\small
\caption{History length at 200k SFT steps, 3-layer SIDs, verbalized history, CPT init. The
variable arm samples lengths in $[5,25]$. Evaluation-set sizes differ because longer
histories restrict the eligible user population.}
\label{tab:history}
\begin{tabular}{lccccr}
\toprule
History & HR@3 & HR@5 & HR@10 & NDCG@10 \\
\midrule
Fixed 10       & \textbf{0.0399} & \textbf{0.0509} & \textbf{0.0667} & \textbf{0.0420} \\
Fixed 50       & 0.0322 & 0.0413 & 0.0544 & 0.0336 \\
Variable 5--25 & 0.0279 & 0.0362 & 0.0492 & 0.0296\\
\bottomrule
\end{tabular}
\end{table}

More history is not better under a fixed training budget. At 200k steps, fixed-10 leads
fixed-50 by 22.6\% relative HR@10 (0.0667 vs.\ 0.0544) and leads the variable-length arm by
35.6\% (0.0492). The ordering is consistent across all four metrics.

\textbf{Why longer histories cost accuracy.} Two mechanisms plausibly contribute. A 50-item
history is roughly five times as many tokens per example, so a fixed step budget buys
proportionally fewer gradient updates per unit of sequence content. And the prediction
target sits further from the most recent --- and most predictive --- engagements in the
attention window.

\textbf{Comparability caveat.} The fixed-50 arm can only be trained and evaluated on users
with at least 50 recorded engagements, a heavier-usage population, and its evaluation set is
correspondingly smaller. Heavier users are not obviously easier or harder to predict --- they
supply more signal but also engage more diffusely --- so we cannot isolate this confound, and
the fixed-50 result should be read as indicative rather than as a controlled comparison. Adding profile features to the fixed-50 arm does not recover
the deficit.

\textbf{Variable-length histories.} Training a single generalist across history lengths is
harder still: at 200k steps the variable arm trails fixed-10 by 35.6\% relative HR@10. The
effect is therefore not specific to length \emph{variability} --- both lengthening histories
and making them heterogeneous cost accuracy at fixed compute.

\textbf{Implications.} The practical reading is not that short histories carry more signal,
but that the specialist is cheaper to train to a given quality --- and, for a deployment,
fixed-length prompts also give predictable serving latency and KV-cache behavior that a
variable-length model does not. The countervailing argument is that one variable-length model
can be prompted at several history lengths to produce different slates for the same user,
offering diversity a specialist cannot without training multiple models.

\subsection{Base Model: Does Capacity Change the Answers?}
\label{sec:exp-model}

\begin{table}[t]
\centering\small
\caption{Base model comparison at 200k SFT steps, 3-layer SIDs, history length 10,
verbalized history.}
\label{tab:base-model}
\begin{tabular}{lccccr}
\toprule
Model & HR@3 & HR@5 & HR@10 & NDCG@10\\
\midrule
Llama-3.2-3B & 0.0399 & \textbf{0.0509} & \textbf{0.0667} & \textbf{0.0420} \\
Gemma-4-12B  & \textbf{0.0400} & 0.0507 & 0.0661 & 0.0418 \\
Qwen3-4B     & 0.0372 & 0.0470 & 0.0610 & 0.0387 \\
\bottomrule
\end{tabular}
\end{table}

Base-model capacity buys strikingly little. At matched 200k steps, Gemma-4-Unified-12B \cite{gemma} and Llama-3.2-3B \cite{llama3} are indistinguishable --- 0.0661 against 0.0667 HR@10, a 0.9\% difference in
Llama's favor --- despite a fourfold difference in parameter count. Qwen3-4B-Instruct-2507\cite{qwen},
though also larger than Llama, trails it by 9.3\% HR@10 and 8.5\% NDCG@10. The ordering does
not track parameter count at all.

Two conclusions follow. First, all three model families reach the same qualitative answers on
SID depth, prompt rendering, and history length, so the design guidance in this section is
not an artifact of one base model. Second, a $4\times$ increase in parameters returns no
measurable accuracy gain, which makes the 3B model the clear operating point once serving
cost enters the comparison. We conjecture that the bottleneck is learning a newly initialized
$3 \times 2048 = 6{,}144$-token vocabulary and its sequential structure --- a task to which
pretrained language ability transfers only weakly --- so that available training steps, not
parameters, is the binding constraint.

\subsection{Reinforcement Learning Post-Training}
\label{sec:exp-rl}

Hit rate and NDCG are ranking objectives that the token-level cross-entropy of SFT
optimizes only indirectly, so post-training the generator against a ranking reward is a
natural next step. We ran 1000 GRPO~\cite{grpo} steps from our best SFT checkpoint under
two reward designs. The first is standard GRPO with an \emph{accuracy} reward: each sampled
completion is scored by whether it matches the ground-truth SID exactly. The second follows
MiniOneRec~\cite{onerec2025} in rolling out with beam search and combining accuracy with a
\emph{ranking} reward, so that a correct item is credited more when it appears higher in
the generated beam --- aligning the training signal more closely with the NDCG we report.

\begin{table}[t]
\centering\small
\caption{GRPO post-training against the SFT checkpoint it initializes from, 3-layer SIDs,
history length 10, 1000 GRPO steps.}
\label{tab:rl}
\begin{tabular}{lccccc}
\toprule
Reward & HR@3 & HR@5 & HR@10 & NDCG@5 & NDCG@10 \\
\midrule
SFT Baseline          & \textbf{0.0399} & \textbf{0.0509} & \textbf{0.0667} & \textbf{0.0366} & \textbf{0.0420} \\
Accuracy                    & 0.0397 & 0.0502 & 0.0651 & 0.0365 & 0.0413 \\
Accuracy $+$ ranking (beam) & 0.0395 & 0.0503 & 0.0651 & 0.0365 & 0.0413 \\
\bottomrule
\end{tabular}
\end{table}

Neither reward improved on SFT (Table~\ref{tab:rl}); both were marginally worse, by 2.4\%
relative HR@10 and 1.7\% NDCG@10. The more informative observation is that the two arms are
indistinguishable from one another --- identical to four decimal places on HR@10, NDCG@5,
and NDCG@10 --- even though the ranking reward was designed specifically to shape the
ordering within the beam and the accuracy reward is indifferent to it. A reward that
targets the metric we care about produced the same model as one that does not, which
suggests the limitation lies upstream of reward design.

We see two reasons, and report them because we believe the failure mode is intrinsic to
generative recommendation in this regime rather than specific to our implementation.

\textbf{The action sequence is too short to benefit from policy optimization.} RL
post-training has been most effective on tasks where the model emits long reasoning traces~\cite{instructgpt,prm},
and where a reward on the final answer can be redistributed across many intermediate token
choices that the policy is free to reorganize. Our completions are $L = 3$ code tokens,
conditioned on a prompt roughly eighty times longer. There is no intermediate structure to
reorganize: the policy makes three categorical decisions, each already directly supervised
by SFT's cross-entropy against the ground-truth code at that position. Sequence-level credit
assignment --- the mechanism through which RL adds value over next-token supervision --- has
almost nothing to assign. The prompt-to-completion ratio also makes rollouts expensive
relative to the signal they carry, since each sample requires a forward pass over a
$\sim$250-token context to produce three tokens of decision.

\textbf{The reward is too sparse to estimate an advantage.} Because our absolute hit rates
are low --- an HR@10 near 0.067 means the correct item appears in a slate of ten roughly one
time in fifteen --- a group of sampled rollouts usually contains no hit at all. Every
completion in the group then receives the same reward, the within-group advantage that GRPO
computes is identically zero, and the group contributes no gradient. Across both runs we
observed this in roughly 95\% of rollout groups: about one group in twenty carried any
learning signal, while all of them incurred full rollout cost. This also explains why the
ranking reward bought nothing. A ranking term can only discriminate among rollouts that
contain a hit to rank, and in the overwhelming majority of groups there was none, so the
richer reward collapsed to the same constant-zero advantage as the binary one. Consistent
with this, mean reward is flat from the first hundred training steps to the last, and the
final KL divergence from initialization is under $6 \times 10^{-3}$ --- the policy barely
moved, because there was almost no gradient to move it.

\textbf{Implications for future work.} We do not recommend RL post-training for generative
recommendation in this regime, but we also do not read our result as evidence that the
direction is unpromising --- rather, that the reward must be densified before policy
optimization has anything to work with. Shaping the reward to give graded semantic credit
looks most promising to us: the coarse-to-fine prefix structure of Section~\ref{sec:sid}
means a generated SID sharing a two-layer prefix with the target is a near miss rather than
a miss, and scoring it accordingly would convert a binary signal into an ordinal one that
varies within almost every group. Alternatives include biasing rollout sampling toward users
whose targets the model can occasionally reach, or enlarging the group size, so that groups
contain reward variance by construction rather than by luck. Each attacks the same
bottleneck: the fraction of rollout groups with non-zero advantage. We suggest reporting
that fraction as a first-class diagnostic in any such attempt, since in our runs it was
already at its final value from the first training step --- long before the downstream
metrics revealed anything.
\section{Conclusion}
\label{sec:conclusion}

We described an end-to-end generative recommendation system for Facebook Groups content, built toward serving Forum, a new standalone application for people heavily invested in Groups. Because Forum has little interaction data of its own, our design rests on two forms of transfer: training on the larger corpus of Groups engagements, and reusing hierarchical, prefix-based semantic IDs learned by an RQ-VAE on cross-platform Facebook Feed content. A 3B-parameter instruction-tuned language model, first grounded in the new SID vocabulary through bidirectional text$\leftrightarrow$SID alignment, learns to generate the SID of a user's next positive interaction directly.Our evaluation isolates four design choices, and the cheap ones matter most. SID depth dominates: the 3-layer configuration outperforms the 4-layer one by 2.2$\times$ HR@10 at matched budget, traded off against a higher collision rate. Verbalizing the engagement history with action verbs gains 2.0\% HR@10 at negligible token cost — more than a full continued-pre-training stage buys, and at a fraction of the cost. Longer and variable-length histories hurt under fixed compute. And base-model capacity buys little: our 3B model and a 12B model are practically indistinguishable (with the 3B model leading by 0.9\% HR@10) despite a fourfold difference in parameters, while a 4B model trails both outright, suggesting the binding constraint is learning a newly initialized 6,144-token vocabulary rather than any deficit in pretrained language ability.Finally, GRPO post-training against an accuracy and ranking based reward did not improve on SFT: with only three or four action tokens per rollout and a hit rate near 0.067, over 95\% of rollout groups contained no reward variance and therefore produced no gradient.

\newpage

\clearpage
\newpage
\bibliographystyle{assets/plainnat}
\bibliography{paper}

\clearpage
\newpage


\end{document}